\documentclass{appolb}
\usepackage{graphicx,amsmath,amssymb,bbm,mathtools}

\usepackage{url}

\newcommand{\B}{{{\mathcal{A}}^{\star}}}
\newcommand{\G}{{\mathcal A}}
\newcommand{\f}{{\mathbf f}}

\newtheorem{proposition}{Proposition}[section]

\begin{document}
\title{An example of a deterministic cellular automaton exhibiting linear-exponential
convergence to the steady state 
}
\author{Henryk Fuk\'s \and Joel Midgley-Volpato
\address{Department of Mathematics and Statistics, Brock University\\
St. Catharines, Ontario L2S 3A1, Canada }
}
\maketitle
\begin{abstract}
In a recent paper \cite{paper55} we presented an example of a 3-state cellular automaton which exhibits behaviour analogous to degenerate hyperbolicity
often observed in finite-dimensional dynamical systems. 
We also calculated densities of 0, 1 and 2 after $n$ iterations of this rule, using finite state machines to conjecture
patterns present in preimage sets. Here we re-derive the same formulae in a rigorous way, without resorting to any semi-empirical methods.
This is done by analyzing the behaviour of continuous clusters of symbols and by considering their interactions. 
\end{abstract}

The general question of finding the iterates of the Bernoulli measure under a given cellular automaton (CA) has been
subject of many recent studies, including, among others, 
\cite{Lind84,Ferrari2000,Maas2003,Host2003,Pivato2002,Pivato2004,Maas2006} and \cite{Maas2006b}.
A more specific question of this type is sometimes called  \emph{the density response problem}:
If the probability of occurence of a certain state in the initial configuration drawn from a Bernoulli distribution is given, what is the probability of occurence
of this state after $n$ iterations of the CA rule?

Of course, one could ask a similar question about the probability of occurence of longer blocks of symbols
after $n$ iterations of the rule. Due to the complexity of CA dynamics, it is clear that questions of this
type are rather hopeless if one wants to know the answer for an arbitrary rule. In spite of this, it may still
be possible to find the answer if the rule is sufficiently simple. 

One of the methods which can be used to do this is studying the structure of preimages of short blocks and detecting patterns present in them.
This approach has been successfully used for a number of deterministic CA rules, such as
elementary  rules 172, 142, 130 (references \cite{paper39},  \cite{paper27}, and \cite{paper40} respectively), and several others.
It has also been used for a special class of probabilistic CA  known as single-transition $\alpha$-asynchronous rules \cite{paper44}. 

Cellular automata are infinitely-dimensional dynamical systems, yet a behaviour similar to hyperbolicity
in finite-dimensional systems has been observed in many of them. In particular, in some
binary  cellular automata in one dimension, known as \emph{asymptotic emulators of identity},   if the initial configuration is drawn from a Bernoulli distribution, 
the expected proportion of ones (or zeros) usually tends to its stationary value exponentially fast \cite{paper52}. This type of behaviour is
quite common in many other dynamical systems. For example, in a linear continuous-time dynamical system given by $\dot{\mathbf{x}}=A\mathbf{x}$, if 
 $\mathbf{x}:\mathbb{R} \to
\mathbb{R}^n$ and $A$ is a real $n \times n$ matrix with all eigenvalues distinct and having negative real parts, $\mathbf{x}(t)$ tends to zero
exponentially fast as $t \to \infty$. Exponential convergence is also observed  in nonlinear systems
$\dot{\mathbf{x}}=\mathbf{f}(\mathbf{x})$ (where $\mathbf{f}:\mathbb{R}^n\to\mathbb{R}^n$)  in the vicinity
of a hyperbolic fixed point, as long as the Jacobian matrix of $\mathbf{f}$ evaluated at the fixed point has only
distinct eigenvalues with negative real parts.

If, on the other hand, the matrix $A$ in  $\dot{\mathbf{x}}=A\mathbf{x}$ has degenerate (repeated) eigenvalues, the convergence to the fixed point can be polynomial-exponential, that is, of the form $P(t)e^{-bt}$, where
$P(t)$ is a polynomial and $b>0$.
Finite dimensional discrete-time dynamical systems can exhibit analogous behaviour. Consider, for example, the linear system 
\begin{equation}
\left[\begin{array}{c}
 x_{n+1}\\
 y_{n+1}
\end{array} \right]=
\left[\begin{array}{rc}
 0 & 1\\
 -\frac{1}{4} & 1
\end{array} \right]
\left[\begin{array}{c}
 x_{n}\\
 y_{n}
\end{array} \right].
\end{equation}
The matrix on the right hand side  has a degenerate (double) eigenvalue~$\frac{1}{2}$, and therefore
the convergence to the fixed point $(0,0)$ is expected to be polynomial-exponential (linear-exponential
in this case). Indeed, if we explicitly solve the above equation for $x_n$ and $y_n$, we obtain
\begin{equation}
\left[\begin{array}{c}
 x_{n}\\
 y_{n}
\end{array} \right]=\left(\frac{1}{2} \right)^n
\left[\begin{array}{cc}
 1-n & 2n\\
 -\frac{n}{2}& 1+n
\end{array} \right]
\left[\begin{array}{c}
 x_{0}\\
 y_{0}
\end{array} \right],
\end{equation}
and we can clearly see the aforementioned linear-exponential convergence.

Very recently, a \textit{probabilistic} CA has been  discovered \cite{probcabookchapter} where the density of ones converges to 
its stationary value in a linear-exponential fashion, just like in the above example of a degenerate hyperbolic fixed point
in a finite-dimensional dynamical systems. This probabilistic CA could be viewed as 
a simple model for diffusion of innovations, spread of rumors, or a similar process involving transport
of information between neighbours. More precisely, it consists of an infinite one-dimensional lattice where each site is occupied by
an individual who has already adopted the innovation (state 1) or who has not adopted it yet (state 0). Once the individual
adopts the innovation, he remains in state 1 forever. Individuals in state 0 can change  their states to 1
(adopt the innovation) with probabilities depending on the state of nearest neighbours. This process can be formally described as a radius 1 binary probabilistic CA with the following
transition probabilities,
\begin{align} \label{adpodef}
 w(1|000)&=0,\, w(1|001)=p,\,w(1|010)=1,\,w(1|011)=1,\\
 w(1|100)&=q,\,w(1|101)=r,\,w(1|110)=1,\,w(1|111)=1, \nonumber
\end{align}
where $p,q,r$ are fixed parameters of the model, $p,q,r \in[0,1]$. By transition probability $w(d|abc)$ one means the
probability that site in state $b$ with neighbours $a$ and $c$ changes its state to $d$ in one time step.
One can show that for a certain choice of parameters  $p,q$ and $r$, the expected value of the density of ones converges to its steady state
in a linear-exponential fashion.

Even more recently,  we found a \textit{deterministic} rule with three states which exhibits
the same kind of behaviour~\cite{paper55}. This rule, to be defined in sec. 2, will be the subject of our subsequent discussion.
In ~\cite{paper55} we studied the structure of preimages of 0, 1 and 2 under the action of this rule, and by employing finite state machines,
we found some patterns in the preimages sets, which in turn allowed us to derive explicit expressions for densities of 0, 1 and 2 after
$n$ iterations. The finite state machines used in the derivation were constructed semi-empirically, and no proof of their correctness was given.
In the current paper we wish to fill this gap and present a more formal derivation of the aforementioned expressions, without 
resorting to finite state machines.

\section{Basic definition}
We will start from some basic definitions. For ${\mathcal{A}}=\{0,1,2\}$,
a finite sequence of elements of ${\mathcal{A}}$, $\mathbf{b}=b_1b_2\ldots, b_{n}$, will be called a \emph{block} 
 (or \emph{word})  of length $n$. The set of all blocks of all possible lengths will be denoted by ${\mathcal{A}}^{\star}$.

Let $f: \mathcal{A}^3 \to \mathcal{A}$ be a local function of a nearest-neighbour cellular automaton.
A {\em block evolution operator} corresponding to  $f$ is a mapping
 $\f:\B \mapsto \B$ defined as follows. Let  $\mathbf{a}=a_1a_2 \ldots a_{n}\in \G^n$
where $n \geq 3$. Then $\f(\mathbf{a})$ is a block of length $n-2$ defined as
\begin{equation}
\f(\mathbf{a}) = f(a_1,a_{2},a_{3})
f(a_2,a_{3},a_{4})\ldots
f(a_{n-2},a_{n-1},a_{n}).
\end{equation}
If  $\mathbf{f}(\mathbf{b})=\mathbf{a}$, than we will say that $\mathbf{b}$ is a preimage of
$\mathbf{a}$, and write $\mathbf{b} \in \mathbf{f}^{-1}(\mathbf{a})$.
Similarly, if $\mathbf{f}^n(\mathbf{b})=\mathbf{a}$, than we will say that $\mathbf{b}$ is an
\emph{$n$-step preimage} of
$\mathbf{a}$, and write $\mathbf{b} \in \mathbf{f}^{-n}(\mathbf{a})$.

Let the \emph{density polynomial} associated with a  string $\mathbf{b}=b_1b_2\ldots b_n$ be defined 
as 
\begin{equation}
 \Psi_{\mathbf{b}}(p,q,r)=p^{\#_0 (\mathbf{b})} q^{\#_1 (\mathbf{b})} r^{\#_2 (\mathbf{b})},
\end{equation}
where $\#_i (\mathbf{b})$ is the number of occurrences of symbol  $i$ in $\mathbf{b}$.
If $A$ is a set of strings, we define the density polynomial associated with $A$ as
\begin{equation}
 \Psi_{A}(p,q,r)=\sum_{\mathbf{a} \in A} \Psi_{\mathbf{a}}(p,q,r).
\end{equation}

One can easily show (in a manner similar as done in \cite{paper52}) that if one starts with a bi-infinite string of symbols drawn from 
a Bernoulli distribution where probabilities of $0,1$ and $2$ are, respectively, $p,q$ and $r$, then
the expected proportion of sites in state $k$  after $n$ iterations of rule $f$
 is given by $\Psi_{\f^{-n}(k)}(p,q,r)$.
This quantity will be called \emph{density} of symbols $k$ after $n$ iterations of $f$.

\section{The local rule and its properties}
Let us now describe the CA rule which will be the subject of this contribution.  While studying properties of various
3-state CA rules, we came across an interesting specimen of a nearest-neighbour (radius 1) rule with a local
function defined as follows,
\begin{equation}
f(x_1,x_2,x_3) =\begin{cases}
x_3 &  \mathrm{for\,\,} x_1=x_2>x_3,\\
x_2 &  \mathrm{otherwise},
\end{cases}
\end{equation} 
where $x_1,x_2,x_3 \in \{0,1,2\}$.
The origins of this rule have been described in~\cite{paper55}. Here we will only note that it can be equivalently
defined as  
\begin{equation}
f(x_1,x_2,x_3) =\begin{cases}
0 &  \mathrm{for\,\,} (x_1,x_2,x_3)=(1,1,0) \mathrm{\,\,or\,\,} (x_1,x_2,x_3)=(2,2,0),\\
1 &  \mathrm{for\,\,} (x_1,x_2,x_3)=(2,2,1),\\
x_2 &  \mathrm{otherwise},
\end{cases}
\end{equation} 
which makes it clear that it differs from the identity rule only on three neighbourhood configurations, $(1,1,0)$, $(2,2,0)$ and $(2,2,1)$.

\begin{figure}[t]
\begin{center}
\includegraphics[scale=0.7]{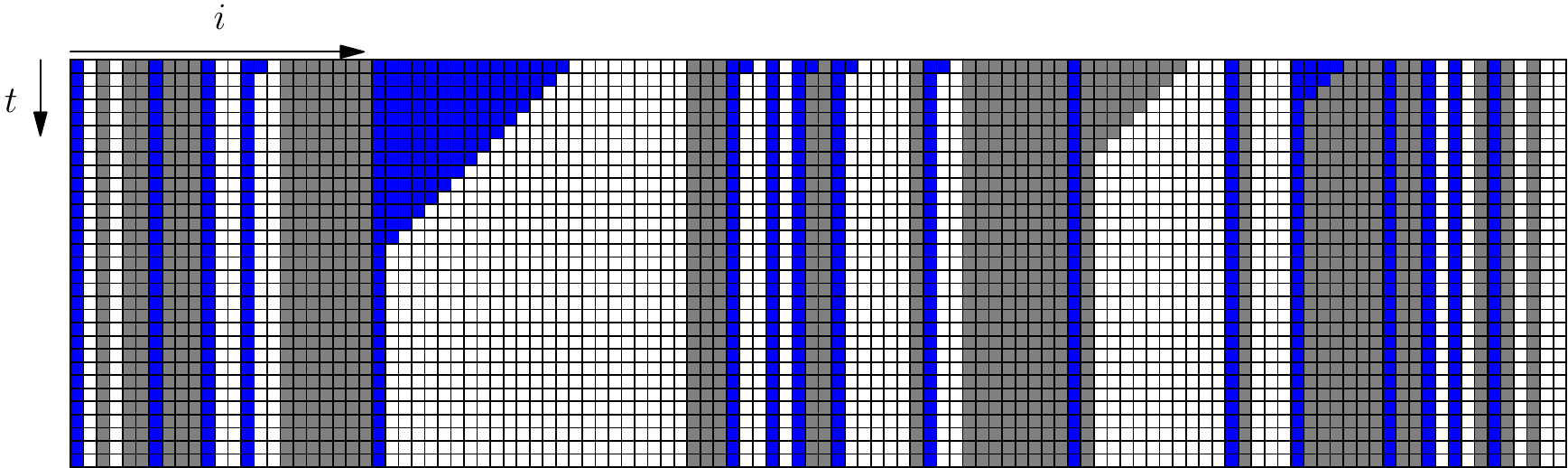}
 \end{center}
\caption{Sample spatio-temporal pattern generated by 3-state rule 140. White, lighter gray and
darker gray cells (blue in color version) correspond, respectively, to 0, 1 and~2.} \label{samplepatern}
\end{figure}

Figure~\ref{samplepatern} shows an example of a  spatio-temporal pattern generated by this rule, using periodic boundary conditions. It
has some important properties which will be relevant to further discussion. First of all, note that $f(x_1,x_2,x_3) \leq x_2$, meaning 
that the state of a given cell cannot increase. This implies that $f(\star,0,\star)=0$, where $\star$ denotes an arbitrary symbol
from the set $\{0,1,2\}$. Zero is thus a quiescent state for this CA.

We also have $f(0,1,\star)=1$ and $f(2,1,\star)=1$, which implies that a site in state 1 located at the beginning of a continuous cluster of 1's
of any length (even of length 1, meaning isolated 1) remain in state 1 forever. The same is true for 2: since $f(0,2,\star)=2$ and $f(1,2,\star)=2$,  a site in state 2 located at the beginning of a continuous cluster of 2's  stays in state 2 forever. This can be observed in Figure~\ref{samplepatern}. In fact, words such as $01$ or $02$
remain unchanged when the rule is iterated, and no information can propagate through a pair sites which are in states $01$ or $02$. We note in passing that 
in the CA theory such words are called \emph{blocking words}, and the rules with blocking words are known to be \emph{almost equicontinuous} \cite{Kurka2009}.

Further inspection of Figure~\ref{samplepatern} reveals that a continuous cluster of zeros grows to the left 
if it is preceded by a cluster of 2's longer than 1, and it also grows to the left if is preceded by a cluster of 1's longer than 1.

\section{Structure of preimages of 1}
We want to find all strings $\mathbf{b}$ of length $2n+1$ such that 
$\f^n(\mathbf{b})=1$.  We know from the definition of the rule that information can propagate only from the right to the left,
thus the first $n-2$ entries of $b$ are arbitrary. We will represent $n$-step preimages of 1 in the form
\begin{equation*}
\mathbf{b}= \underbrace{\star \star \ldots \star}_{n-2}  a_1 a_2 \mathbf{a}_3  c_1c_2 \ldots c_{n},
 \end{equation*}
where the allowed values of  $a_1,a_2, a_3$  (to be called a \textit{prefix}) and $c_1c_2 \ldots c_{n}$ (to be called a \textit{postfix}) need to be determined.
The central site of the preimage string will be, as in the above, denoted by a bold symbol.
Since our CA rule has three states, there are $3^3=27$ possible values for the prefix $a_1a_2a_3$. Not all of them are possible,
however. Prefixes $00\mathbf{0}$, $01\mathbf{0}$, $02\mathbf{0}$, $10\mathbf{0}$, $11\mathbf{0}$, $12\mathbf{0}$, $20\mathbf{0}$, $21\mathbf{0}$ and 
$22\mathbf{0}$, which can be represented as $\star \star \mathbf{0}$,
cannot occur in any preimage of 1. This is because $f(\star,0,\star)=0$, meaning that if the central site  is in state 0, it will remain in state 0 forever, and consequently 
 $\f^n(b)=0$ for any string $b$ containing one of the above prefixes.

Moreover, prefixes of the type $\star0\mathbf{2}$ or $\star 1\mathbf{2}$ are not allowed either.  This is because for the central site to become
1, as required, the transition $f(2,2,1)=1$ would have to happen somewhere along the way, and for this the central 2 would have to get 2 as the left neighbour.
Since states can only decrease, not increase, and the left neighbour of the central site is 0 or 1, this is not possible.

By excluding 9 prefixes of the type $\star \star \mathbf{0}$ and 6 prefixes of the type $\star0\mathbf{2}$ or $\star 1\mathbf{2}$ we are left with 12 possibilities, 
$00\mathbf{1}$,
$01\mathbf{1}$,
$02\mathbf{1}$,
$10\mathbf{1}$,
$11\mathbf{1}$,
$12\mathbf{1}$,
$20\mathbf{1}$,
$21\mathbf{1}$,
$22\mathbf{1}$,
$02\mathbf{2}$,
$12\mathbf{2}$,
$22\mathbf{2}$.
All of them are allowed in preimages of 1, providing that an appropriate suffix is added. In what follows we will find conditions which these suffices
need to satisfy. We will divide the possible prefixes into four different types (the reason for this will soon become clear):
\begin{enumerate}
 \item $\star 1 \mathbf{1}$, $22\mathbf{1}$
 \item $\star 0\mathbf{1}$, $02\mathbf{1}$, $12\mathbf{1}$
 \item $02\mathbf{2}$, $12\mathbf{2}$
 \item $22\mathbf{2}$
\end{enumerate}
For prefixes of type 1,  $\star 1 \mathbf{1}$ and $22\mathbf{1}$, the central site is in the state 1 already, thus we have to make
sure that it stays in the same state after $n$ iterations of the rule. The left neighbour of the central 1 is 1 (for $\star 1 \mathbf{1}$) or
it will become 1 after one iteration (for $2 2 \mathbf{1}$), thus we could potentially be in a danger of the transition $f(1,1,0)=0$.
This could happen only if the central 1 belongs to a continuous cluster of ones which is followed by 0 -- in such a case, the cluster
of ones will shrink one symbol per time step and the transition $f(1,1,0)=0$ will eventually change the central 1 into 0.
We have, therefore, two choices  in avoiding this scenario: either the central 1 belongs to a cluster of ones followed by 2, which
prevents its shrinking due to the fact that $f(1,1,2)=1$, or it belongs to a cluster of ones which extends all the way to the right.
In other words, the postfix for type 2 must be  of the form 
\begin{equation*}
c_1c_2\ldots c_n=  1^i2 \underbrace{\star \star \ldots \star}_{n-1-i}
\mathrm{\,\,\,\,\,or\,\,\,\,\,}
c_1c_2\ldots c_n=  1^n,
\end{equation*}
where $i\in \{0, 1, \ldots n-1\}$. In the above and in what follows, $1^n$ denotes the symbol 1 repeated $n$ times. We will use
this convention in the rest of the paper.

For type 2, $\star 0\mathbf{1}$, $02\mathbf{1}$, and $12\mathbf{1}$, note that the central site could become 0 only by utilizing
the transition $f(1,1,0)=0$, and for this the left neighbour of the central 1 would have to be 1. This is clearly impossible for 
$\star 0\mathbf{1}$ as the left neighbour 0 will always remain 0. In the case of $02\mathbf{1}$ and $12\mathbf{1}$, the left neighbour
is 2, and it could potentially become 0 by transition $f(2,2,0)=0$. For this, however, we would need
the second left neighbour of the central 1 to be in state 2, but this is impossible because site values never increase. Thus all prefixes
of type 2 belong to preimages of 1, regardless of the suffix.

For type 3, $02\mathbf{2}$, $12\mathbf{2}$, the central site is in state 2, and its left neighbour is guaranteed to be in state 2 forever.
Consequently, the central 2  must change to 1 at some iteration via the $f(2,2,1)=1$ transition, and in order for this to happen,  we need to have
the right neighbour of the central site in state 1 at some point of time. This can happen when the central 2 belongs to a continuous cluster of
2's followed by 1, meaning that the postfix must be of the form
\begin{equation*}
c_1c_2\ldots c_n= 2^i 1 \underbrace{\star \star \ldots \star}_{n-1-i},
\end{equation*}
where $i\in \{0, 1, \ldots n-1\}$.

Type 4 with the prefix  $22\mathbf{2}$ is the most complicated one. Similarly as before, the central site must change from 2 to 1, and this can 
only happen via the transition $f(2,2,1)=1$, but we do not have a guarantee that the left neighbour of the central 2 remains in state 2 forever,
as it was the the case of type 3. Nevertheless, the necessary condition for the suffix is  the same as for type 3, meaning that
the central 2 must belong to a continuous cluster of 2's followed by 1. This in not sufficient, however, because what follows is also important.
In order to understand this clearly, consider two strings of length 13 iterated 6 times: 
\begin{center}
\begin{verbatim}
       0000222221012                   0000122221012
        00022221101                     00012221101
         002221100                       001221100
          0221100                         0121100
           21100                           12100
            100                             210
             0                               1
\end{verbatim}
\end{center}
The first of these strings has prefix 222, and the second one has prefix 122 (thus belonging to type 3). In both of them the central 2
 belongs to a continuous cluster of 2's followed by 1, but the first one does not produce 1 after 6 iterations. This is because the zero
which follows propagates to the left and eventually makes the central site to change to 0 via the transition $f(1,1,0)=0$. Any string containing 222
as a prefix must, therefore, satisfy an additional property preventing zeros to propagate to the left. This can be done by a making the last part of the 
suffix to have the same structure as prefix of type 3, so that the entire suffix takes the form
\begin{equation}
c_1c_2 \ldots c_n=2^i 1 d_1d_2\ldots d_{n-1-i},
\end{equation}
where $i\in \{0, 1, \ldots n-1\}$ and where
\begin{equation}
 d_1d_2\ldots d_{n-1-i} =  1^j2 \underbrace{\star \star \ldots \star}_{n-2-i-j}
\mathrm{\,\,\,\,\,or\,\,\,\,\,} d_1d_2\ldots d_{n-1-i}=1^{n-1-i},
\end{equation} with $j\in \{0,2, \ldots, n-2-i$\}.

Finally, for prefix $222$ there is one more possibility not covered by the above discussion, namely
$c_1c_2 \ldots c_n= 2^{n-2} 1 0$. Below we summarize all these findings in a form of a single proposition.
\begin{proposition}
 Block $b$ belongs to $\f^{-n}(1)$ if and only if it is one of the following four types.
\begin{itemize}
 \item[Type 1:] \begin{equation}
\mathbf{b}= \underbrace{\star \star \ldots \star}_{n-2}  a_1 a_2 a_3 1^i2 \underbrace{\star \star \ldots \star}_{n-1-i}
\mathrm{\,\,\,\,\,or\,\,\,\,\,}
\mathbf{b}= \underbrace{\star \star \ldots \star}_{n-2}  a_1 a_2 a_3 1^n,
\end{equation}
where $a_1a_2a_3\in \{011,111,211,221\}$, $i\in \{0, 1, \ldots n-1\}$;
\item[Type 2:] \begin{equation}
\mathbf{b}= \underbrace{\star \star \ldots \star}_{n-2}  a_1 a_2 a_3 \underbrace{\star \star \ldots \star}_{n},
\end{equation}
where where $a_1a_2a_3\in \{001,101,201,121,021\}$;

\item[Type 3:] \begin{equation}
\mathbf{b}= \underbrace{\star \star \ldots \star}_{n-2}  a_1 a_2 a_3 2^i 1 \underbrace{\star \star \ldots \star}_{n-1-i},
\end{equation}
where where $a_1a_2a_3\in \{022,122\}$, $i\in \{0, 1, \ldots n-1\}$

\item[Type 4a:] \begin{equation}
\mathbf{b}= \underbrace{\star \star \ldots \star}_{n-2}  a_1 a_2 a_3 2^i 1 c_1c_2\ldots c_{n-1-i},
\end{equation}
where  $a_1a_2a_3=222$, $i\in \{0, 1, \ldots n-1\}$ and where
\begin{equation}
 c_1c_2\ldots c_{n-1-i} =  1^j2 \underbrace{\star \star \ldots \star}_{n-2-i-j}
\mathrm{\,\,\,\,\,or\,\,\,\,\,} c_1c_2\ldots c_{n-1-i}=1^{n-1-i}
\end{equation} with $j\in \{0,2, \ldots, n-2-i$\}.

\item[Type 4b:] \begin{equation}
\mathbf{b}= \underbrace{\star \star \ldots \star}_{n-2}  a_1 a_2 a_3 2^{n-2} 1 0,
\end{equation}
where  $a_1a_2a_3=222$.
\end{itemize}
\end{proposition}

\section{Density polynomials for preimages of 1}
Let us now denote the set of strings of type 1 by $T_1$, type 2 by $T_2$ etc., and let us
define $\lambda=p+q+r$.
Density polynomial for $T_1$ will be given by
\begin{multline}
 \Psi_{T_1}(p,q,r)=\sum_{i=0}^{n-1}\lambda^{n-2}(pq^2+q^3+rq^2+r^2q)q^ir \lambda^{n-i-1}\\+
\lambda^{n-2}(pq^2+q^3+rq^2+r^2q) q^n,
\end{multline}
which simplifies to
\begin{equation}
 \Psi_{T_1}(p,q,r)=\lambda^{2n-3} r (\lambda q^2+r^2q) \sum_{i=0}^{n-1}\lambda^{-i}q^i +
\lambda^{n-2}(\lambda q^2+r^2q) q^n.
\end{equation}
By performing summation of the partial geometric sequence in the above, one obtains
\begin{equation}
 \Psi_{T_1}(p,q,r)=\lambda^{n-2} r (\lambda q^2+r^2q) \frac{\lambda^n-q^n}{p+r}+
\lambda^{n-2}(\lambda q^2+r^2q) q^n,
\end{equation}
which further simplifies to
\begin{equation}
  \Psi_{T_1}(p,q,r)= r (\lambda q^2+r^2q) \frac{\lambda^{2n-2}}{p+r} + \lambda^{n-2}  (\lambda q^2+r^2q)\frac{pq^n}{p+r}.
\end{equation}
Similar calculations (omitted here) yield
\begin{equation}
 \Psi_{T_2}(p,q,r)=(\lambda pq + q^2r  +pqr) \lambda^{2n-2},
\end{equation}
\begin{equation}
 \Psi_{T_3}(p,q,r)=\lambda^{n-2} r^2 q (\lambda^n-r^n).
\end{equation}
The type 4a is the most complicated. Let us first compute the density polynomial for the set of strings 
of the form
\begin{equation}
 c_1c_2\ldots c_{k} =  1^j 2 \underbrace{\star \star \ldots \star}_{k-1-j}
\mathrm{\,\,\,\,\,or\,\,\,\,\,} c_1c_2\ldots c_{k}=1^{k}
\end{equation} with $j\in \{0,1, \ldots, k-1$\}. The density polynomial for the above, to be denoted by $h_k(p,q,r)$,
is given by
\begin{equation}
  h_k(p,q,r) = \sum_{j=0}^{k-1} q^j r \lambda^{k-1-j}+q^n= r \frac{\lambda^k-q^k}{p+r} + q^k
=\frac{r \lambda^k}{p+r}+ \frac{pq^k}{p+r}.
\end{equation}
Having this result, we can write the density polynomial for the entire set $T_{4a}$,
\begin{multline}
 \Psi_{T_{4a}}(p,q,r)=\lambda^{n-2} r^3 \sum_{i=0}^{n-1} r^iq h_{n-1-i}(p,q,r)\\
=\lambda^{n-2} r^3 \sum_{i=0}^{n-1} r^iq \frac{r \lambda^{n-1-i}}{p+r}
+ \lambda^{n-2} r^3 \sum_{i=0}^{n-1} r^iq\frac{pq^{n-1-i}}{p+r}\\
=\frac{\lambda^{n-2}q r^4}{p+r} \sum_{i=0}^{n-1} r^i  \lambda^{n-1-i}
+ \frac{\lambda^{n-2} p q r^3}{p+r} \sum_{i=0}^{n-1} r^i q^{n-1-i}\\
=\frac{\lambda^{n-2}q r^4(\lambda^n-r^n)}{(p+r)(p+q)} 
+ \frac{\lambda^{n-2} p q r^3}{p+r} \sum_{i=0}^{n-1} r^i q^{n-1-i}.
\end{multline}
When $q\neq r$, we thus obtain
\begin{equation}
 \Psi_{T_{4a}}(p,q,r)=\frac{\lambda^{n-2}q r^4(\lambda^n-r^n)}{(p+r)(p+q)} 
+ \frac{\lambda^{n-2} p q r^3(q^n-r^n)}{(p+r)(q-r)}.
\end{equation}
When $q=r$, the last sum becomes $\sum_{i=0}^{n-1} r^i q^{n-1-i}=\sum_{i=0}^{n-1} q^{n-1}=q^{n-1}n$, therefore
\begin{equation}
 \Psi_{T_{4a}}(p,q,q)=\frac{\lambda^{n-2}q^5(\lambda^n-q^n)}{(p+q)^2} 
+ \frac{\lambda^{n-2} p  q^3}{p+q}  n q^{n}.
\end{equation}
Finally, type 4b is straightforward,
\begin{equation}
 \Psi_{T_{4b}}(p,q,r)=\lambda^{n-2}r^{n+1} p q.
\end{equation}

We are now ready to compute the density polynomial of preimages of 1, by summing density polynomials for $T_1,T_2,T_3, T_{4a}$ and $T_{4b}$.
This yields, for $q \neq r$,
\begin{multline}
\Psi_{\f^{-n}(1)}(p,q,r)= 
r (\lambda q^2+r^2q) \frac{\lambda^{2n-2}}{p+r} + \lambda^{n-2}  (\lambda q^2+r^2q)\frac{pq^n}{p+r}\\
+(\lambda pq + q^2r  +pqr) \lambda^{2n-2}
+\lambda^{n-2} r^2 q (\lambda^n-r^n)\\
+\frac{\lambda^{n-2}q r^4(\lambda^n-r^n)}{(p+r)(p+q)} 
+ \frac{\lambda^{n-2} p q r^3(q^n-r^n)}{(p+r)(q-r)} + \lambda^{n-2}r^{n+1} p q.
\end{multline}
Collecting together terms for $(q \lambda)^n$, $(r \lambda)^n$, and $\lambda ^{2n}$ we obtain, after some algebra,
\begin{multline} \label{normalpreim1}
\Psi_{\f^{-n}(1)}(p,q,r)= {\frac {  p {q}^{2} \left( -pr+pq+{q}^{2} \right)
\left( q \lambda \right)^{n} }{ \lambda^{2} \left( p+r \right)  \left( q
-r \right) }}\\
+{\frac {q r \left( -{p}^{2}r+{p}^{2}q+p{q}^{2}-2\,pqr+{r}^
{3}-{q}^{2}r \right)   \left( r \lambda \right)^{n}}{ 
\lambda^{2} \left( p+q \right)  \left( q-r \right) }}\\
+{\frac {q
 \left( {p}^{3}+{p}^{2}q+2\,{p}^{2}r+p{r}^{2}+3\,pqr+{r}^{3}+{r}^{2}q+
{q}^{2}r \right)  \lambda^{2n} }
 { \lambda  \left( p+r \right)  \left( p+q
 \right) }},
\end{multline}
which is the same formula as derived in \cite{paper55}.

Similarly, for $q=r$, we obtain
\begin{multline}
\Psi_{\f^{-n}(1)}(p,q,q)= 
q (\lambda q^2+q^3) \frac{\lambda^{2n-2}}{p+q} + \lambda^{n-2}  (\lambda q^2+q^3)\frac{pq^n}{p+q}\\
+(\lambda pq + q^3  +pq^2) \lambda^{2n-2}
+\lambda^{n-2} q^3  (\lambda^n-q^n)\\
\frac{\lambda^{n-2}q^5(\lambda^n-q^n)}{(p+q)^2} 
+ \frac{\lambda^{n-2} p  q^3}{p+q}  n q^{n} +\lambda^{n-2}q^{n+1} p q.
\end{multline}
After simplification and reordering of terms this yields
\begin{multline} \label{P1degenerate}
\Psi_{\f^{-n}(1)}(p,q,q)= {\frac {p{q}^{3} \left( n+1 \right)  \left( q \lambda \right) ^{n}
}{ \lambda^{2} \left( q+p \right) }}
+{\frac {{q}^{2}
 \left( 2\,{p}^{3}+4\,{p}^{2}q+p{q}^{2}-2\,{q}^{3} \right)  \left( q \lambda \right) ^{n}}{ \left( q+p \right) ^{2}  \lambda^{2}}}\\
+{\frac { \left( {p}^{3}+3\,{p}^{2}q+4\,p{q}^{2}+3\,{q}
^{3} \right) q \lambda^{2n}}{ 
 \lambda   \left( q+p \right) ^{2}}},
\end{multline}
which, again, agrees with the result ``guessed'' in \cite{paper55} using finite state machines.

\section{Preimages of 2 and their density polynomials}
From the definition of the rule we know that a site can be in state 2 only if it was in that state at the beginning,
that is, sites in state 0 or 1 cannot change to 2. Moreover, $f(2,2,0)=0$, $f(2,2,1)=1$, and in all other cases $f(a_1, 2, a_3)=2$. This means that a site in state 2
remains in that state forever if it is preceded by 0 or 1. Therefore, any string of the form 
\begin{equation*}
\mathbf{b}= \underbrace{\star \star \ldots \star}_{n-1}0\textbf{2}  \underbrace{\star \star \ldots \star}_{n} \mathrm{\,\,\,\,\,or\,\,\,\,\,}
\mathbf{b}= \underbrace{\star \star \ldots \star}_{n-1}1\textbf{2}  \underbrace{\star \star \ldots \star}_{n}
\end{equation*}
will be an $n$-step preimage of 2, $\f^n(b)=2$.

What if 2 is preceded by 2? In this case it must be followed by a sufficient number of 2's before the first 0 or 1 appears, as
any 0 or 1 at the end of a cluster of 2's shortens such cluster by one on each iteration. Therefore, for $n$ iterations we need $n$ 2's.
We thus need, in order for $\f^n(b)=2$ to hold in this case, 
\begin{equation*}
 \mathbf{b}= \underbrace{\star \star \ldots \star}_{n-1} 2\textbf{2} 2^{n}.
\end{equation*}
The above observations can be summarized as follows.
\begin{proposition}
 Block $b$ belongs to $\f^{-n}(2)$ if and only if it is one of the following three types:
\begin{enumerate}
 \item  $\mathbf{b}= \underbrace{\star \star \ldots \star}_{n-1}02  \underbrace{\star \star \ldots \star}_{n}$,
 \item   $\mathbf{b}= \underbrace{\star \star \ldots \star}_{n-1}12  \underbrace{\star \star \ldots \star}_{n}$,
 \item   $\mathbf{b}= \underbrace{\star \star \ldots \star}_{n-1} 2^{n+2}$.
\end{enumerate}
\end{proposition}
This yields the density polynomial
\begin{equation}
 \Psi_{\f^{-n}(2)}(p,q,r)= (p+q)r\lambda^{2n-1}+ \lambda^{n-1} r^{n+2}.
\end{equation}
\section{Preimages of 0 and their density polynomials}
Since everything what is not a preimage of 1 or 2 must be a preimage of 0,
we have
\begin{equation}
 \Psi_{\f^{-n}(0)}(p,q,r) = \lambda^{2n+1}-\Psi_{\f^{-n}(1)}(p,q,r) -\Psi_{\f^{-n}(2)}(p,q,r).
\end{equation}
After simplification, this yields, for $r \neq q$,
\begin{multline} \label{normalpreim0}
\Psi_{\f^{-n}(0)}(p,q,r)= {\frac { \left( -pr+pq+{q}^{2} \right) p{q}^{2} \left(q \lambda
 \right) ^{n}}{ \lambda^{2} \left( p+r \right)  \left( 
r-q \right) }} \\
+{
\frac {p r \left( -{r}^{2}p+{q}^{2}p+{q}^{3}-{r}^{3}-q{r}^{2} \right) 
 \left( r \lambda \right) ^{n}}{ \lambda^{2} \left( 
p+q \right)  \left( r-q \right) }}
\\
+{\frac { \left( {p}^{3}+2\,{p}^{2}q+2\,{p}^{2}r+2\,{r}^
{2}p+3\,qpr+2\,{q}^{2}p+{r}^{3}+2\,q{r}^{2}+{q}^{3}+{q}^{2}r \right) p
 \lambda^{2n}}{
 \left( p+q \right)  \left( p+r \right)  \lambda }},
\end{multline}
 and for $r=q$,
\begin{multline} \label{P0degenerate}
\Psi_{\f^{-n}(0)}(p,q,q) ={\frac { \left( {p}^{3}+4\,{p}^{2}q+7\,{q}^{2}p+5\,{q}^{3} \right) p
 \lambda^{2n}}{ \lambda 
 \left( p+q \right) ^{2}}}
-{\frac {p{q}^{3} \left( n+1 \right) 
 \left( q \lambda \right) ^{n}}{  \lambda^{2}
 \left( p+q \right) }}
\\
-{\frac {{q}^{2}p \left( 3\,{p}^{2}+8\,pq+6\,{q}
^{2} \right)  \left( q \lambda \right) ^{n}}{ \left( p+q \right) ^
{2} \lambda  ^{2}}}.
\end{multline}
Again, similarly as in the density polynomial for 1, the linear-exponential dependence of the form $( n+1) ( q \lambda )^n$ is present in the second term.

\section{Density of ones}

As already stated, density polynomials $\Psi_{\f^{-n}(k)}(p,q,r)$ represent probability of occurence
of $k$ after $n$ iterations starting from a Bernoulli distribution with probabilities of 0, 1 and 2 equal to, respectively, 
$p$, $q$, and $r$, where $p+q+r=1$. If one starts with a symmetric Bernoulli distribution
where $r=q$, the probability of occurence of 1 after $n$ steps, to be denoted by $P_n(1)$, 
will be given by eq. (\ref{P1degenerate}) as long as one  substitutes $r=q$ and $q=(1-p)/2$. This yields, after simplification,
\begin{equation}
P_n(1)=P_{\infty}(1)-(An+B)\left(\frac{1-p}{2} \right) ^{n},
\end{equation}
where
\begin{align}
A&= \frac{\left( p-1 \right) ^{2}}{4 \left( 1+p
 \right) ^{2}} \, \left( {p}^{3}-p \right), \\
B&= \frac{\left( p-1 \right) ^{2}}{4 \left( 1+p
 \right) ^{2}} \, \left( -{p}^{3}-5\,p-3\,{p}^{2}+1 \right), \\ 
 P_{\infty}(1) &= {\frac { \left( 1-p \right)  \left( {p}^{3}+5\,{p}^{2}-p+3
 \right) }{4 \left( 1+p \right) ^{2}}}.
\end{align}
One can see that for $0<p<1$, $P_n(1)$ tends to  $P_{\infty}(1)$ as $n \to \infty$, and that the convergence is linear-exponential 
in $n$. Such ``degenerate'' convergence takes place for  probability of occurence of 0 as well, as seen in
eq. (\ref{P0degenerate}). 

On the other hand, when $r \neq q $ in the initial Bernoulli distribution, the convergence
is purely exponential, as in eq. (\ref{normalpreim1}) and (\ref{normalpreim0}). 

\section{Conclusions}
We presented an example of a 3-state rule  exhibiting, under certain conditions,  linear-exponential convergence to the steady state. This
phenomenon is remarkably similar to degenerate hyperbolicity in finite dimensional dynamical systems. It is not clear, however,
what is the origin of this analogy. One can speculate that ``simple'' CA rules, such as those which are equicontinuous or almost
equicontinuous, can be somewhat approximated by finite-dimensional systems. Local structure theory could possibly be applicable 
in this case, as it allows to construct finite-dimensional systems approximating orbits of Bernoulli measure under the action of a 
given CA. It is hoped that this contribution inspires further research on this subject.

\section*{Acknowledgments}
The author acknowledges financial support from the Natural Sciences and
Engineering Research Council of Canada (NSERC) in the form of Discovery Grant.
This work was made possible by the facilities of the Shared
Hierarchical Academic Research Computing Network (SHARCNET:
www.sharcnet.ca) and
Compute/Calcul Canada.


\begin{thebibliography}{10}

\bibitem{paper55}
H.~Fuk\'s and J.~Midgley-Volpato, ``An example of degenerate hyperbolicity in
  cellular automaton with 3 states,'' in {\em 21st International Workshop on
  Cellular Automata and Discrete Complex Systems}, J.~Kari, I.~T\"orm\"a, and
  M.~Szabados, eds., vol.~24 of {\em TUCS Lecture Notes}, pp.~47--55.
\newblock Turku, Finland, 2015.
\newblock \href{http://xxx.lanl.gov/abs/arXiv:1506.06649}{{\tt
  arXiv:1506.06649}}.

\bibitem{Lind84}
D.~A. Lind, ``Applications of ergodic theory and sofic systems to cellular
  automata,'' {\em Phys. D} {\bf 10} (1984), no.~1-2, 36--44. Cellular automata
  (Los Alamos, N.M., 1983).

\bibitem{Ferrari2000}
P.~A. Ferrari, A.~Maass, S.~Mart{\'{\i}}nez, and P.~Ney, ``Ces\`aro mean
  distribution of group automata starting from measures with summable decay,''
  {\em Ergodic Theory Dynam. Systems} {\bf 20} (2000), no.~6, 1657--1670.

\bibitem{Maas2003}
A.~Maass and S.~Martinez, ``Evolution of probability measures by cellular
  automata on algebraic topological markov chains,'' {\em Biol. Res.} {\bf 36}
  (2003), no.~1, 113--118.

\bibitem{Host2003}
B.~Host, A.~Maass, and S.~Mart{\'{\i}}nez, ``Uniform {B}ernoulli measure in
  dynamics of permutative cellular automata with algebraic local rules,'' {\em
  Discrete Contin. Dyn. Syst.} {\bf 9} (2003), no.~6, 1423--1446.

\bibitem{Pivato2002}
M.~Pivato and R.~Yassawi, ``Limit measures for affine cellular automata,'' {\em
  Ergodic Theory Dynam. Systems} {\bf 22} (2002), no.~4, 1269--1287.

\bibitem{Pivato2004}
M.~Pivato and R.~Yassawi, ``Limit measures for affine cellular automata.
  {II},'' {\em Ergodic Theory Dynam. Systems} {\bf 24} (2004), no.~6,
  1961--1980.

\bibitem{Maas2006}
A.~{Maass}, S.~{Mart{\'{\i}}nez}, and M.~{Sobottka}, ``Limit measures for
  affine cellular automata on topological markov subgroups,'' {\em
  Nonlinearity} {\bf 19} (Sept., 2006) 2137--2147.

\bibitem{Maas2006b}
A.~Maass, S.~Martinez, M.~Pivato, and R.~Yassawi, ``Asymptotic randomization of
  subgroup shifts by linear cellular automata,'' {\em Ergodic Theory and
  Dynamical Systems} {\bf 26} (2006), no.~04, 1203--1224.

\bibitem{paper39}
H.~Fuk{\'s}, ``Probabilistic initial value problem for cellular automaton rule
  172,'' {\em DMTCS proc.} {\bf AL} (2010) 31--44,
  \href{http://xxx.lanl.gov/abs/arXiv:1007.1026}{{\tt arXiv:1007.1026}}.

\bibitem{paper27}
H.~Fuk{\'s}, ``Dynamics of the cellular automaton rule 142,'' {\em Complex
  Systems} {\bf 16} (2006) 123--138.

\bibitem{paper40}
H.~Fuk\'s and A.~Skelton, ``Response curves for cellular automata in one and
  two dimensions -- an example of rigorous calculations,'' {\em International
  Journal of Natural Computing Research} {\bf 1} (2010) 85--99,
  \href{http://xxx.lanl.gov/abs/arXiv:1108.1987}{{\tt arXiv:1108.1987}}.

\bibitem{paper44}
H.~Fuk\'s and A.~Skelton, ``Orbits of {B}ernoulli measure in asynchronous
  cellular automata,'' {\em Dis. Math. Theor. Comp. Science} {\bf AP} (2011)
  95--112.

\bibitem{paper52}
H.~Fuk\'s and J.-M.~G. Soto, ``Exponential convergence to equilibrium in
  cellular automata asymptotically emulating identity,'' {\em Complex Systems}
  {\bf 23} (2014) 1--26, \href{http://xxx.lanl.gov/abs/arXiv:1306.1189}{{\tt
  arXiv:1306.1189}}.

\bibitem{probcabookchapter}
H.~Fuk\'s, ``An example of computation of the density of ones in probabilistic
  cellular automata by direct recursion.'' submitted for publication, 2015.

\bibitem{Kurka2009}
P.~K{\r{u}}rka, ``Topological dynamics of cellular automata,'' in {\em
  Encyclopedia of Complexity and System Science}, R.~A. Meyers, ed.
\newblock Springer, 2009.

\end{thebibliography}
\providecommand{\href}[2]{#2}\begingroup\raggedright\endgroup

\end{document}